\documentclass[pra,superscriptaddress,twocolumn,amsmath,amssymb]{revtex4}
\usepackage{etex}
\usepackage{amsfonts}
\usepackage{amssymb}
\usepackage{mathtools}
\newtheorem{thm}{Theorem}

\usepackage{graphicx}
\usepackage{subfigure}
\usepackage{epstopdf}
\usepackage{mathrsfs}
\usepackage{color,soul}
\usepackage[normalem]{ulem}
\usepackage{natbib}
\usepackage{bbold}
\usepackage{placeins}
\usepackage{braket}
\usepackage{soul,xcolor}
\usepackage{epstopdf}
\usepackage{tabu}
\usepackage{array}
\usepackage{bm}
\usepackage{upgreek}
\usepackage{multirow}
\usepackage[utf8]{inputenc}
\usepackage[colorlinks = truelinkcolor = blue, urlcolor  = blue, citecolor = blue, anchorcolor = blue]{hyperref}
\hypersetup{
	colorlinks   = true, 
	urlcolor     = blue, 
	linkcolor    = blue, 
	citecolor   = blue 
}

\newcommand{\opt}{\operatorname}
\makeatletter

\begin{document} 	
	\title{Projective measurements under qubit quantum channels}
	
		\author{Javid Naikoo}
	\email{j.naikoo@cent.uw.edu.pl}
	\affiliation{Centre for Quantum Optical Technologies, Centre of New Technologies, University of
		Warsaw, Banacha 2c, 02-097 Warsaw, Poland}

	\author{Subhashish Banerjee}
	\email{subhashish@iitj.ac.in}
	\affiliation{Indian Institute of Technology Jodhpur, Jodhpur 342011, India}

	\author{A. K. Pan}
	\email{akp@nitp.ac.in}
	\affiliation{National Institute of Technology Patna, Ashok Rajpath, Patna, Bihar 800005, India }
	
	\author{Sibasish Ghosh}
	\email{sibasish@imsc.res.in}
	\affiliation{Optics \& Quantum Information Group,The Institute of Mathematical Sciences, HBNI,CIT Campus, Taramani, Chennai - 600113, India }
	
	\begin{abstract}
		\noindent 
			      The action of qubit channels on projective measurements on a qubit state is used to establish an equivalence between channels and properties of generalized  measurements characterized by \textit{bias} and \textit{sharpness} parameters. This can be interpreted as shifting the description of measurement dynamics from the Schrodinger to the Heisenberg picture. In particular,  unital quantum channels are shown to induce \textit{unbiased} measurements. The Markovian channels are found to be equivalent to measurements for which sharpness is a monotonically decreasing function of time. These results are illustrated by considering various noise channels. Further, the effect of \textit{bias} and \textit{sharpness} parameters on the energy cost of  a measurement and its interplay with  non-Markovianity of  dynamics is also discussed.
	\end{abstract}
	
	\maketitle

	\section{Introduction}
	
	Measurement in quantum theory plays a crucial role compared to its classical counterpart. The ineluctable feature of any quantum measurement is that it entails an interaction between the measuring apparatus so that the 
	observed system necessarily gets  entangled with the system of observing apparatus  \cite{busch1996quantum,braginsky1995quantum}. The text-book description of quantum measurement usually deals with the ideal measurement when there is an one-to-one correspondence between system and apparatus states is achieved. Such an ideal measurement scenario can be modelled by a complete set of orthogonal projectors on the given Hilbert space of the system. However, in practice, the system-apparatus correspondence may not be achieved in general measurement scenario. In such a case, the orthogonal projectors are needed to be replaced by the 
	so-called positive operator-valued measures (POVMs) which is a set of Hermitian, positive semi-definite operators commonly denoted as $\{ E_i\}$,  that sum up to identity, i.e., $\sum_{i = 1}^{n} E_i = \mathbb{1}$. If all the elements of a POVM are projective, i.e., $E_i = \Pi_{i}=\ket{\phi_i}\bra{\phi_i}$, where $\{ \ket{\phi_i}\}$ is an orthonormal basis, then the measurement is called \textit{sharp}.  There are pertinent schemes where the generalized measurement involving POVMs outperform the projective measurements, such as,  quantum tomography \cite{braginsky1995quantum}, unambiguous state discrimination \cite{busch2010coexistence}, quantum cryptography \cite{yu2010joint,busch2010unsharp,das2018testing}, device-independent randomness certification \cite{acin16} and many more.\bigskip

	Mathematically, a  two-outcome POVMs in two dimension in its general form can be written as \cite{busch2010coexistence,stano2008,yu2010joint}
	\begin{equation}\label{eq:Epm}
		E_{\pm}(x, \vec{m}) =  \frac{\mathbb{1} \pm (\mathbb{1} x + \vec{m} \cdot \vec{\sigma})}{2}. 
	\end{equation} 
	where $x$ and $|\vec{m}|$ are called \textit{bias} and \textit{sharpness} parameters respectively. The positivity of the POVMs $E_{\pm}(x, \vec{m})$ demands that the following condition be satisfied
	\begin{equation}
		|x| + |\vec{m}| \le 1.
	\end{equation} 
	For ideal sharp measurement scenario, $|x|=0$ and  $|\vec{m}|=1$. The notion of \textit{bias} and \textit{shapness} capture the deviation from the ideal projective measurements, but arises due to different physical reasons. The \textit{sharpness} parameter is linked with the precision of measurement arising due to operational indistinguishability between the probability distributions corresponding to the post-measurement apparatus states and thus  $|\vec{m}|=1$ implies vanishingly small overlap between them. On the other hand \textit{bias} parameter quantifies the tendency of a measurement to favor one state over the other.  When $|x|=0$ the POVM $E_{\pm}(0, \vec{m})$ is called \textit{unbiased}, meaning that the outcomes of measurement are purely random if the system is prepared in a maximally mixed state, i.e., $\opt{Tr} \{ E_{+}(0, \vec{m}) \mathbb{1}/2\} = \opt{Tr} \{ E_{-}(0, \vec{m}) \mathbb{1}/2\} = 1/2$.\bigskip

	The POVM elements in Eq. (\ref{eq:Epm}), can be viewed as an \textit{affine} transformation on a pure state $\rho = \frac{1}{2}( \mathbb{1} + \vec{r} \cdot \sigma)$, with $\vec{r} = (\sin\theta \cos\phi, \sin\theta \sin\phi, \cos\theta)$ being the Bloch vector. The post-measurement state becomes $\rho^\pm = \frac{1}{2}( \mathbb{1} + \vec{s}^{\pm} \cdot \sigma)$, such that 
	\begin{equation}
	\vec{s}^{\pm} = A^{\pm}\vec{r} + \vec{T}^{\pm}.
	\end{equation}
	Here, $A_{ij}^{\pm} = \frac{1}{2} \opt{Tr} \{\sigma_i E_{\pm} [\sigma_j]\}$ and $T_{i}^{\pm} = \frac{1}{2} \opt{Tr} \{\sigma_i E_{\pm}[\mathbb{1}]\}$, with $E_{j}[ \bm{\omega}] =  E_{j} \bm{\omega} E_{j}^\dagger$, $j=\pm$. We have 
	\begin{align}
		A_{ii}^{\pm} &= \frac{1}{4} \big[ (1\pm x)^2 + 2m_i^2 - |\vec{m}|^2 \big],\\\nonumber 
		A_{ij}^{\pm} &=\frac{m_i m_j}{2}, ~~{\rm for}~~i\ne j ;\ \ 
		T_{i}^{\pm} = \frac{m_i}{2}(1 + x), 
	\end{align}
	with $i,j = 1,2,3$ and $\pm$ pertaining to the POVM element $E_{\pm}$ being used. \bigskip

	The effect of \textit{biased}-\textit{unsharp} measurements on various quantum correlations has been studied in \cite{busch2010unsharp,das2018testing,swati2017probing}.
	 The \textit{biased} and \textit{unsharp} measurements have been realized experimentally using quantum feedback stabilization of photon number in a cavity \cite{sayrin2011real}. The \textit{unsharp} measurements were studied with qubit observables through an Arthur–Kelly-type joint measurement model for qubits \cite{pal2011approximate}. Recently, implementation of generalized measurements on a qudit via quantum walks was  also proposed \cite{Zhiao2019implement}.\bigskip

	 The quantum dynamics, in its idealized version without presence of environment, is governed by unitary evolution. In realistic scenario, the most general evolution is governed by quantum channels which are characterized by suitably formulated Kraus operators. The action of a quantum channel is conveniently studied in the Schr\"{o}dinger picture, while its effect on the operator, leading to the operators subsequent evolution, requires the use of the Heisenberg picture.    In this work, we address the following issues. Under the action of a quantum dynamical process ({\it e.g.}, a quantum channel), how does an ideal projector evolve, and whether such a process transform it to a POVM. In particular, we examine the types of quantum channels that leads to the two-outcome \textit{biased}-\textit{unsharp} POVMs.  Moreover, we have investigated the behavior of the \textit{bias} and \textit{sharpness} parameters under the effect of Markovian as well as non-Markovian nature of the quantum dynamics. Further, in a direction towards motivating our study, we look at the comparison of energy costs for implementing such measurements under different kinds of open system dynamics. \bigskip

	The work is organized as follows: In Sec. (\ref{UnitalvsBUM}) we start with a brief review  of quantum channels and discuss the effect of dynamics on ideal measurements. The arguments are made rigorous by proving two theorems establishing that the  conjugate of a unital (non-unital) channel  generates \textit{unbiased} (\textit{biased}) POVM. Also, the conjugate of a Markovian channel is shown to lead to \textit{unsharp}  measurements such that the \textit{sharpness} is a monotonically decreasing function of time. The effect  of \textit{bias} and \textit{sharpness}, in the context of non-Markovian dynamics, on the energy cost of measurements is discussed in Sec. (\ref{sec:ECost}).  We conclude in Sec. (\ref{Conclusion}).
	
	\section{Quantum channels and \textit{biased}-\textit{unsharp} POVMs }\label{UnitalvsBUM}
	Mathematically, \textit{quantum channels} are linear maps $\mathcal{E}: \mathcal{S}(X) \rightarrow \mathcal{S}(Y)$,  such that $ \mathcal{S}(X)$ ($ \mathcal{S}(Y)$)  is the set of all density operators acting on $X$ ($Y$) \cite{watrous2018theory}.  Geometrically, the quantum channel $\mathcal{E}$ is an \textit{affine} transformation \cite{ruskai2002analysis}. An elegant description of quantum channels is given in terms of \textit{operator sum representation}, such that an initial density matrix $\rho$ is evolved to some final density matrix $\rho^\prime$
	\begin{equation}
		\rho^\prime = \mathcal{E} [\rho]  = \sum\limits_{i} \mathcal{K}_i \rho  \mathcal{K}_i^\dagger.
	\end{equation}
	Here $\mathcal{K}_i$ are the Kraus operators and satisfy the completeness condition $\sum_i \mathcal{K}_i^\dagger \mathcal{K}_i = \mathbb{1}$.    The conjugate channel $\mathcal{E}^\dagger$ corresponding to $\mathcal{E}$ is defined such that $\mathcal{E}^\dagger[\rho] = \sum\limits_{i} \mathcal{K}_i^\dagger \rho  \mathcal{K}_i$.  
	
	An important class of quantum channels are the \textit{unital channels}, each of which maps Identity operator to itself, that is, $\mathcal{U}[\mathbb{1}] = \mathbb{1}$. Examples include the phase damping, depolarizing and Pauli channels \cite{sbrg,omkar}. A typical example of non-unital channel is the amplitude damping channel \cite{sbbook,sgad,sbgp}.  Note that the conjugate channel of each quantum channel is unital. For unital qubit channels, the following properties are equivalent \cite{mendl2009unital}:
	\begin{enumerate}
		\item $\mathcal{E}$ is unital if $\mathcal{E}[\mathbb{1}] = \mathbb{1}$.
		\item 
		$\mathcal{E}$ can be realized as a random unitary map: ${\cal E}(\rho) = \sum_{i} p_iU_i\rho U_i^{\dagger}$ with $U_i$'s being unitary and $p_i$'s being probabilities such that $\sum_i p_i = 1$.
	\end{enumerate}

	The projective measurements are mapped to the POVMs due to the action of channels. As an example, consider a  unital qubit channel $T$ which  can be described as \cite{mendl2009unital} 	$T (\rho) = \sum_{i}^{} \lambda_i U^\dagger_i \rho U_i = \rho^\prime$,	with $U_iU_i^\dagger = \mathbb{1}$, and $ \sum_{i}^{} \lambda_i =1$.  Let a qubit projective measurement be denoted by $\Pi^{\pm}$ such that  the probability of obtaining the outcome $\pm 1$ is given by  $\rm{prob}({\pm}1) = \opt{Tr}\{ \Pi^{\pm} \rho^\prime \} = \opt{Tr} \{\Pi^{\pm} \sum_{i}^{} \lambda_i U^\dagger_i \rho U_i  \} =\opt{Tr} \{ \sum_{i}^{}  \lambda_i U_i \Pi^{\pm} U^\dagger_i \rho \} = \opt{Tr} \{ E_{\pm} \rho \}$.  	Here $E_{\pm} =  \sum_{i}^{}  \lambda_i U_i \Pi^{\pm} U^\dagger_i$ can be identified as the POVM elements, in the sense that the projectors evolve to POVMs: $E_{\pm} \ge 0$ and $E_+ + E_- = I$, through the dynamics.  Since $\Pi^{\pm}$ is a projector having unit trace,  therefore $\opt{Tr} \{U_i \Pi^{\pm} U^\dagger_i\} = 1 $, as $U_i$ is trace preserving. This indicates that $	\opt{Tr} \{E_{\pm}\} =  \opt{Tr}\{\sum_{i}^{}  \lambda_i U_i \Pi^{\pm} U^\dagger_i \} = \sum_{i}^{}  \lambda_i \opt{Tr} \{U_i\Pi^{\pm} U^\dagger_i\},
	= \sum_{i}^{}  \lambda_i =1$.  But, the trace of  POVM element $ E_{\pm}$ in Eq. (\ref{eq:Epm}) is  $	\opt{Tr} \{E_{\pm}\} = \opt{Tr} \{\frac{\mathbb{1} \pm (\mathbb{1} x + \vec{m}.\vec{\sigma})}{2} \} = 1 \pm x $.   \textit{Hence a unital qubit channel acting on projectors leads to \textit{unbiased} POVMs}. \bigskip

	\begin{table*}[htp]
		\centering
		\caption{\ttfamily  The \textit{bias} and \textit{sharpness} parameters corresponding to different noise channels. Here, $\Theta$ is the measurement parameter defined in Eq. (\ref{measurement}).
			Detailed discussion about these channels  (including their time dependence) can be found in the cited references. 
		}
		\begin{tabular}{ |p{3cm}|p{6cm}|p{3cm}|p{4.2cm}|}
			\hline
			Channel   (unital/non-unital)                                      & Kraus operators & \textit{Bias}  & \textit{Sharpness} \\
			\hline
			Random Telegraph Noise (RTN) \cite{daffer2004depolarizing},  unital, with memory            &  $R_1 = \sqrt{\frac{1+\Lambda(t)}{2}} \mathbb{1}$, $R_2 =\sqrt{\frac{1-\Lambda(t)}{2}} \sigma_z$                 &            0     &   $\sqrt{\cos^2\Theta + \Lambda^2(t) \sin^2\Theta}$\\
			\hline
			Phase Damping (PD) \cite{nielsen2002quantum}, unital, without memory  & \small $ P_0 = \begin{pmatrix}
				1 &  0 \\
				0  &  \sqrt{1-\lambda}
			\end{pmatrix}$, $P_1 = \begin{pmatrix}
				1 & 0\\
				0 & \sqrt{\lambda}
			\end{pmatrix}$. \normalsize & 0           &   $\sqrt{1- \lambda \sin^2\Theta}$          \\
			\hline
			Depolarizing \cite{nielsen2002quantum}, unital, without memory  & \small $ D_0 = \sqrt{1-q}~ \mathbb{1}$ \normalsize, \small $D_i = \sqrt{q/3}~ \sigma_i$ $i=1,2,3$ \normalsize & 0           &   $ 1 - 4q/3$          \\
			\hline
			Amplitude Damping (AD) \cite{nielsen2002quantum},  non-unital, without memory  &  $A_0 = \small \begin{pmatrix} 1 & 0\\ 0 & \sqrt{\gamma} \end{pmatrix}$, $A_1 = \begin{pmatrix} 0 & \sqrt{1-\gamma}\\ 0 &0   \end{pmatrix}$ \normalsize  & $|\gamma \cos\Theta|$    & $\sqrt{1-\gamma} ~\sin\Theta$  \\
			\hline
			AD \cite{bylicka2014non}, non-unital,  with memory  &  $\tilde{A}_0 = \small \begin{pmatrix} 1 & 0\\ 0 & G(t) \end{pmatrix}$, $\tilde{A}_1 = \begin{pmatrix} 0 & \sqrt{1-|G(t)|^2}\\ 0 &0   \end{pmatrix}$ \normalsize  & $|G^2(t) \cos\Theta -1|$    & $|G(t)| \sqrt{\sin^2\Theta + |G(t)|^2 \cos^2\Theta}$  \\
			\hline
			Generalized AD (GAD) \cite{nielsen2002quantum}, non-unital, without memory  & $G_0 = \small  \begin{pmatrix} \sqrt{p} & 0\\ 0 & \sqrt{p(1- \gamma)} \end{pmatrix}$,\newline \small $G_1 =  \begin{pmatrix} 0 & \sqrt{p \gamma}\\ 0 & 0 \end{pmatrix}$\normalsize, 
			\newline	\small $G_2 = \begin{pmatrix} \sqrt{(1-p) (1- \gamma)} & 0\\ 0 & 1  \end{pmatrix}$ \normalsize, \newline \small $G_3 =  \begin{pmatrix} 0 & 0\\ \sqrt{(1-p) \gamma} & 0  \end{pmatrix}$.\normalsize  \normalsize&  $ |(2p-1) \gamma \cos\Theta|$ & \small $\sqrt{(1-\gamma)(1-\gamma \sin^2\Theta)}$\normalsize   \\
			\hline
		\end{tabular}\label{tabBiasUnsharp}
	\end{table*}
	
	We provide an illustrative example to demonstrate the interplay of the \textit{bias} and \textit{sharpness} parameters with the nature of the underlying dynamics. For this let us assume a qubit  interacting with random telegraph noise \cite{daffer2004depolarizing}, characterized by the stochastic variable  $\Gamma(t)$ switching at a rate $\gamma$ between $\pm 1$. The variable $\Gamma(t)$ satisfies  the correlation    $\langle \Gamma (t)  \Gamma(s) \rangle = a^{2} e^{-(t-s)/\tau}$, where  $a$ is the qubit-RTN coupling strength, and $\tau = \frac{1}{2 \gamma}$.   The reduced   dynamics of qubit  is  governed  by following Kraus operators
	\begin{equation}\label{KrausRTN}
		R_1 (\nu) = \sqrt{\frac{1+ \Lambda(\nu)}{2}} \mathbb{1}, \qquad  R_2 (\nu) = \sqrt{\frac{1- \Lambda(\nu)}{2}} \sigma_z.
	\end{equation}
	Here, $\Lambda(\nu) = e^{-\nu } \big[\cos(\mu \nu) + \frac{\sin(\mu \nu)}{\mu}\big]$ is the memory kernel with $\mu = \sqrt{(4 a \tau)^2 - 1}$ and $\nu=\frac{t}{2\tau}=\gamma t$ is  a dimensionless parameter.  When $0 \le 4a \tau < 1$, the dynamics is  damped  with the frequency parameter $\mu$ imaginary with magnitude less than unity. At $4a \tau =1$, the memory kernel $\Lambda = e^{\nu} (1-\nu)$, which is unity at the initial time and approaches zero as time approaches infinity. For $4a \tau > 1$, the dynamics exhibits damped harmonic oscillations in the interval $[-1,1]$. The former and later scenarios correspond to the Markovian and non-Markovian dynamics, respectively \cite{naikoo2019facets,pradeep,qds}.\bigskip

	Let the initial system be represented by a pure qubit state (as it will turn out, the conclusion we draw are actually independent of the initial state)
	\begin{equation}\label{rho}
		\rho = \begin{pmatrix}
			\cos^2(\frac{\theta}{2})                          &   \frac{1}{2} e^{-i \phi} \sin(\theta)  \\\\
			\frac{1}{2} e^{i \phi} \sin(\theta)          &   \sin^2(\frac{\theta}{2})
		\end{pmatrix}.
	\end{equation}
	Here, $0 \le \theta \le \pi$ and $0 \le \phi \le 2 \pi$. Further, we use the general dichotomic observable $\hat{\mathcal{Q}}$  parametrized as 
	\begin{equation}\label{measurement}
		\hat{\mathcal{Q}} = \begin{pmatrix}
			\cos\Theta     &   e^{i \Phi} \sin \Theta\\
			e^{-i\Phi} \sin\Theta &   -\cos\Theta
		\end{pmatrix},
	\end{equation}
	with $0 \le \Theta < \pi$ and $0\le \Phi \le 2 \pi$ \cite{murnaghan1962unitary}. The corresponding eigen projectors are $\Pi^{\pm} =\frac{1}{2}[ \mathbb{1} \pm \hat{\mathcal{Q}}]$. The expectation values of $\Pi^{\pm}$ under RTN dynamics is given by
	\begin{equation}
		\langle \Pi^{\pm} \rangle = \opt{Tr} \{ \Pi^{\pm} \sum_{i=1}^2 R_i \rho R_{i}^{\dagger}  \} =  \opt{Tr} \{ \rho  \sum_{i=1}^2 R_{i}^{\dagger} \Pi^{\pm} R_i  \}.
	\end{equation}
	We can then identify the term $ \sum_{i=1}^2 R_{i}^{\dagger} \Pi^{\pm} R_i $ as the POVM $E^{\pm}$, which when compared to Eq. (\ref{eq:Epm}), gives the \textit{bias} and \textit{shaprness} parameters. 
	Thus the action of a noisy channel on the dynamics of the qubit can be viewed as a generalized measurement.
	Equating $\opt{Tr} \{  \rho \sum_{i=1}^{2}  R^\dagger_i \Pi^{\pm} R_i \}$ with $ \opt{Tr} \{E_{\pm}(x, \vec{m})  \rho  \} $, one can obtain the \textit{bias} and sharpness parameters as
	\begin{equation}
		x= 0, \quad {\rm and}\quad |\vec{m}| = \sqrt{\cos^2\Theta + \Lambda^2(t) \sin^2\Theta}.
	\end{equation}
	Thus the evolution of projector under this channel provides the \textit{unbiased} POVMs and the \textit{sharpness} parameter contains the \textit{memory kernel} $\Lambda(t)$, which in turn decides whether the dynamics is Markovian or non-Markovian.  Since RTN constitutes a unital channel, this is in accordance with our above considerations about unital channels inducing \textit{unbiased} POVMs. Similarly, one finds that for the amplitude damping channel with memory, see Table (\ref{tabBiasUnsharp}), the \textit{memory kernel} $G(t)$, is present both in \textit{bias} as well as  \textit{sharpness} parameter. \bigskip

	 Before proceeding further we provide a brief introduction of  the \textit{Mueller matrix} formulation  which will be used to prove two general results.\bigskip

	A linear operation $\mathcal{E}$ and its adjoint $\mathcal{E}^\dagger$ are defined in terms of Hilbert-Schmidt inner product $\langle \rho \sigma \rangle = \opt{Tr}\{\rho^\dagger \sigma \}$, such that $\opt{Tr} \{ [\mathcal{E}(\rho)]^\dagger \sigma \} = \opt{Tr} \{ \rho^\dagger \mathcal{E}^\dagger (\sigma) \}$, with the Kraus operators of $\mathcal{E}^\dagger$ being the adjoint of those of  $\mathcal{E}$ \cite{ruskai2002analysis}. Further,  $\mathcal{E}$  is trace preserving if and only if $\mathcal{E}^\dagger$ is unital.  Now, the (linear) action of a qubit channel on the four dimensional column vector $(1, r_x, r_y, r_z)^T$ to produce the four dimensional 
	column vector $(1, s_x, s_y, s_z)^T$ is obtained by a $4 \times 4$ real matrix ( say $M$). In the Optics literature, $M$ is generally called a Mueller matrix. Here $(r_x, r_y, r_z) ~[(s_x, s_y, s_z)] $ is the Bloch vector of the input [output] qubit state ${\rho}_{in} = (1/2)(I + r_x{\sigma}_x + r_y{\sigma}_y + r_z{\sigma}_z)~ [{\rho}_{out} = (1/2)(I + 
	s_x{\sigma}_x + s_y{\sigma}_y + s_z{\sigma}_z)].$ \bigskip

	The qubit channel $\mathcal{E}$ - a $4 \times  4$ matrix with complex  entries in general) - acting on the column vector $((1 +  r_z)/2, (r_x - ir_y)/2, (r_x + ir_y)/2, (1 - r_z)/2)^T$ of the input state ${\rho}_{in}$, produces another column vector $((1 + s_z)/2, (s_x - is_y)/2, (s_x + is_y)/2, (1 - s_z)/2)^T$ of the output state  ${\rho}_{out}$. This transformation matrix is the Mueller matrix $M$ under conjugation, i.e.,  every entry of $\mathcal{E}$ is a linear combination of the  entries of $M$ and vice-versa. The coefficients of these linear combinations are  independent of the parameters of the input as well as output qubit states. Trace-preservation condition of the channel $\mathcal{E}$ demands that the 1st row  of the Mueller matrix M must be: $(1, 0, 0, 0)$, i.e., $M = \begin{pmatrix}  1  & \mathbf{0} \\  \mathbf{t}  &  \mathbf{\Lambda} \end{pmatrix}$,  with $\mathbf{\Lambda}$ a $3 \times 3$  real matrix and ${\bf 0} = (0, 0, 0)$ and ${\bf t} = (t_1, t_2, t_3)^T$ are real vectors.  The map $\mathcal{E}$ is unital if and only if $\mathbf{t}=0$.  It can be verified easily that the Mueller matrix corresponding to the conjugate channel ${\cal E}^{\dagger}$ of the qubit channel ${\cal E}$ is given by:  $M_{{\cal E}^{\dagger}} = M_{{\cal E}}^T$, the transpose of the Mueller matrix for ${\cal E}$. \bigskip

	While finding out the canonical form of a qubit channel, it is useful  to work with the Mueller matrix $M$ rather than the channel matrix $\mathcal{E}$. Thus, for example, for any two $2 \times 2$ special unitary matrices $U$ and $V$, the qubit state $ V \mathcal{E} (U{\rho}_{in}U^{\dagger})V^{\dagger} $ corresponds to the action of the Mueller matrix $(1 \oplus R_V)M  (1 \oplus R_U)$, where $R_U (R_V)$ is the $3 \times 3$ real rotation matrix corresponding to $U (V)$. Using this fact and the idea of singular value decomposition, one can now make the last $3 \times 3$ block sub-matrix of $M$ to be a real diagonal matrix: $diag({\lambda}_1, {\lambda}_2, {\lambda}_3)$. Thus a canonical $M$ matrix is represented by six real parameters (satisfying the CPTP condition) -- three  $t_1, t_2, t_3$ -- say, corresponding to the 1st column vector $(1, t_1, t_2, t_3)^T$ of $M$ and the aforesaid remaining three parameters ${\lambda}_1, {\lambda}_2$, and ${\lambda}_3$. For unital channels $ t_1 = t_2 = t_3 = 0$ \cite{ruskai2002analysis}. \bigskip

	We now prove two theorems based on the observations at  the beginning of this section.

	\begin{thm}\label{thm:unital}
		The conjugate of a unital (non-unital) qubit channel generates an \textit{unbiased} (\textit{biased}) POVM.
	\end{thm} 
	
	\textit{Proof}: The case of unital channels has already been discussed earlier in this section. Here we provide the explicit form of the POVM after the effect of the dual channel.
	Consider a unital channel $\mathcal{E}[\rho] = \sum_j p_j U_j \rho U_j^\dagger$, with $0 \le p_j \le 1$, $\sum_j p_j = 1$, and $U_j U_j^\dagger = U_j^\dagger U_j = \mathbb{1}$.  The action of a channel on a state $\rho$ is equivalent to the action of its \textit{conjugate channel} on the operator, $\opt{Tr}\{ \mathcal{E}[\rho] \Pi^{\pm}\} = \opt{Tr}\{ \sum_j p_j A_j \rho A_j^\dagger \Pi^{\pm} \} = \opt{Tr} \{\rho \sum_j p_j A_j^\dagger \Pi^{\pm} A_j  \} = \opt{Tr}\{\rho \mathcal{E}^\dagger[\Pi^{\pm}] \}$. Therefore, a  projective measurement $\mathcal{M} = \{\Pi^{\pm} = \frac{1}{2}(\mathbb{1} \pm \hat{m}.\vec{\sigma})\}$, under the action of a conjugate of \textit{unital} channel  $\mathcal{E}$, evolves as $\mathcal{M} = \mathcal{E}^\dagger [\mathcal{M}]  = \{ \mathcal{E}^\dagger [\Pi^+], \mathcal{E}^\dagger[\Pi^-]\}$, such that 
	\begin{align}
		\mathcal{E}^\dagger[\Pi^{\pm}] &= \sum_j p_j U_j^\dagger \Pi^{\pm} U_j = \sum_j p_j U_j^\dagger (\frac{\mathbb{1} \pm \hat{m} \cdot \vec{\sigma}}{2}) U_j \nonumber \\
		&= \frac{1}{2} \Big[ \mathbb{1} \pm \Big[\big(\sum_j p_j R_{U_j^\dagger}\big) \hat{m}\Big] \cdot \vec{\sigma} \Big].
	\end{align}
	The effect on the Bloch vector is a series of rotations $R_{U^\dagger_j} [\cdot] = U^\dagger_j \cdot U_j$ weighted by the probability $p_j$. It follows that $\mathcal{E}^\dagger[\Pi^+] + \mathcal{E}^\dagger[\Pi^-] = \mathbb{1}$ and forms a POVM.\bigskip

	Let us now consider the case when the channel $\mathcal{E}$ is  \textit{non-unital}, i.e., $\mathcal{E}[\mathbb{1}] = \sum_j p_j  A_j \mathbb{1} A_j^\dagger \neq \mathbb{1}$, $0 \le p_j \le 1$, $\sum_j p_j =1$, and $\sum_j p_j A_j^\dagger A_j = \mathbb{1}$. The last condition ensures that the conjugate channel is unital $\mathcal{E}^\dagger[\mathbb{1}] = \sum_j p_j A_j^\dagger \mathbb{1} A_j  = \sum_j p_j A_j^\dagger A_j = \mathbb{1}$. The action of a quantum channel and its conjugate on an input $R = (r_0~ r_1~ r_2~ r_3)^T$, $T$ being the transposition operation, can be described by  \textit{Mueller} matrices, as discussed above, with the following representation
	\begin{equation}
		M_{\mathcal{E}} = \begin{pmatrix}
			1    &      0           &        0         &  0\\
			t_1  & \lambda_1  &        0         &  0 \\
			t_2  &     0           &  \lambda_2  &  0 \\
			t_3  &    0            &        0         &    \lambda_3
		\end{pmatrix}, ~{\rm and }~  M_{\mathcal{E}^\dagger} = \begin{pmatrix}
			1    &      t_1           &        t_2         &   t_3\\
			0    & \lambda_1  &        0         &  0 \\
			0    &     0           &  \lambda_2  &  0 \\
			0    &    0            &        0         &    \lambda_3
		\end{pmatrix},
	\end{equation}
	respectively. It immediately follows that under
	\begin{align}
		M_{\mathcal{E}} &: \mathbb{1}\rightarrow \mathbb{1} + \vec{t} \cdot \vec{\sigma}; ~~ \sigma_j \rightarrow  \lambda_j \sigma_j,\nonumber \\
		{\rm and} ~~~~ M_{\mathcal{E}^\dagger} &:  \mathbb{1}\rightarrow \mathbb{1}; ~~ \sigma_j \rightarrow t_j \mathbb{1}+ \lambda_j \sigma_j,
	\end{align}
	with $j=1,2,3$. Hence it follows that under the action of a unital channel, the output of the Muller matrices for the channel as well as its adjoint are equal. The action on the corresponding inputs translate to 
	\begin{align}\label{eq:Mueller}
		M_{\mathcal{E}} [\rho]&= M_{\mathcal{E}} [\frac{1}{2} (\mathbb{1} + \vec{m} \cdot\vec{\sigma})]  =   \frac{1}{2} [\mathbb{1} + (\vec{t} + \vec{m}^{\prime} )\cdot \vec{\sigma}],   \nonumber \\
		M_{\mathcal{E}^\dagger} [\Pi^{\pm}] &= M_{\mathcal{E}^\dagger} [\frac{1}{2} (\mathbb{1} \pm \hat{m} \cdot \vec{\sigma})] =   \frac{1}{2} [ (1 \pm x )\mathbb{1} \pm \vec{m}^\prime \cdot \vec{\sigma}].
	\end{align}
	Thus the \textit{bias} and \textit{sharpness} parameters are identified with $x = \hat{m}\cdot \vec{t}$, and  $\vec{m}^\prime = (m_1 \lambda_1, ~ m_2 \lambda_2, ~ m_3 \lambda_3)^T$, respectively. Therefore the resulting POVM is \textit{unbiased} if $\vec{t}=0$, i.e., if the channel is unital. Further, $\opt{Tr} M_{\mathcal{E}^\dagger} [\Pi^{\pm}] = 1 \pm x$, implies that the trace preserving conjugate channels must be \textit{unbiased}.   It  follows that the  conjugate of a unital qubit channel acting on a projective measurement generates an \textit{unbiased} POVM. \hfill$\blacksquare$ \bigskip

	
	\begin{thm}\label{thm:Markov}
		Under the action of the dual of any qubit Markovian dynamics, a projective measurement gets mapped into  a POVM for which the sharpness parameter decreases monotonically with time. 
	\end{thm}

	\textit{Proof}:
	We make use of the fact that under Markovian dynamics ${\cal E}_{\tau}$, the trace distance (TD)  between two arbitrary states $\rho_1$ and $\rho_2$ is a monotonically decreasing function of time. Consider the case when $\rho(0) = \frac{1}{2}(\mathbb{1} + \vec{n} \cdot \sigma)$ and $\sigma(0) = \frac{1}{2} \mathbb{1}$. The action of map $\mathcal{E}$ is described by the corresponding Mueller matrix as shown in Eq. (\ref{eq:Mueller}). We have
	\begin{align}
		{\rm TD} &= \frac{1}{2} \big|\big| \mathcal{E}_{\uptau} [\rho(0)] -  \mathcal{E}_{\uptau}[\sigma(0)]  \big|\big|_1 \nonumber \\
		&= \frac{1}{2} \big|\big| \mathcal{E}_{\uptau} [\frac{1}{2}(\mathbb{1} + \vec{n} \cdot \vec{\sigma})] -  \mathcal{E}_{\uptau}[\frac{1}{2} \mathbb{1}]  \big|\big|_1 \nonumber \\
		&= \big|\big| \frac{1}{2} [\mathbb{1} + \vec{t}(\uptau) \cdot \vec{\sigma}  + \sum_j n_j \lambda_j(\uptau) \sigma_j]  -  \frac{1}{2}[\mathbb{1} + \vec{t}(\uptau) \cdot \vec{\sigma}] \big|\big|_1 \nonumber \\
		&= \alpha \big| \vec{n}^\prime (\uptau) \big|.
	\end{align}
	Here, $\alpha = \frac{1}{2}|| \hat{n}^\prime (\uptau) \cdot \vec{\sigma} ||_1 = \frac{1}{2} || \frac{1}{2} (\mathbb{1} + \hat{n}^\prime \cdot \sigma) - \frac{1}{2} (\mathbb{1} - \hat{n}^\prime \cdot \sigma) ||_1$ is a constant and $\vec{n}^\prime (\uptau) = \big(n_1 \lambda_1(\uptau), n_2 \lambda_2(\uptau), n_3 \lambda_3(\uptau) \big)$. Therefore, for ${\rm TD}$ to be a monotonically decreasing function, $| \vec{n}^\prime (\uptau) |$ must monotonically decrease and saturate to zero when time $\uptau \rightarrow \infty$. This, in turn implies that $\lambda_j (\uptau)$ is a decreasing function and converges to zero in the limit $\uptau \rightarrow \infty$.\bigskip

	From the statements made below Eq. (14), the sharpness parameter   $| \vec{m}^\prime(\uptau) | = \sqrt{\sum_j |m_j \lambda_j (\uptau)|^2}$. Since $\lambda_j(\uptau)$ is monotonically decreasing as shown above, we conclude that, for Markovian dynamics, the \textit{shaprness} parameter $| \vec{m}^\prime(\uptau) |$ must also be a monotonically decreasing function of time and should saturate to zero as $ \uptau \rightarrow \infty$. As a consequence of Theorem 2, the non-monotonic behavior in time of the sharpness parameter would be an indicator of P-indivisible form of non-Markovianity  \cite{blpv}. \hfill$\blacksquare$   \bigskip

	An example of such a scenario would be furnished by the RTN noise channel discussed above.

	\section{Effect on energy cost of a measurement}\label{sec:ECost}
	In this section,  we discuss the energy requirement for performing a general quantum measurement \cite{abdelkhalek2016fundamental}. The physical implementation of a measuring device  comprises of two steps: a \textit{measurement step} which consists of storing a particular measurement outcome in a register $M$, and a \textit{resetting step}, which resets the measuring device to its initial state for  repeated implementation \cite{abdelkhalek2016fundamental,deleter}. The total energy cost of the complete measurement process amount to the energy costs in these two steps together. \bigskip

	\begin{figure}[ht!]
		\centering
		\includegraphics[width=80mm]{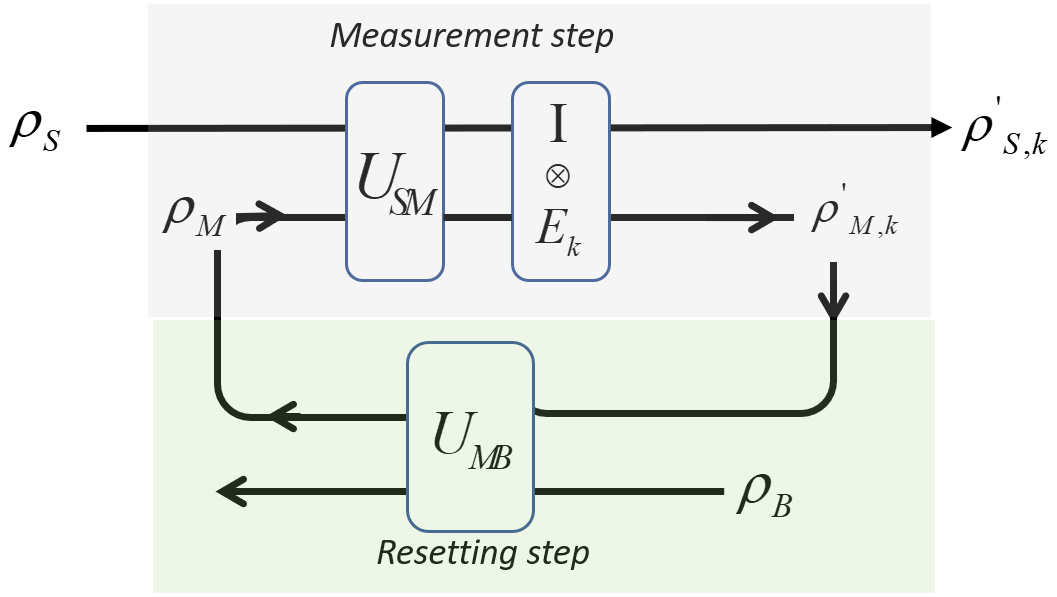}
		\caption{(Color online) Measurement scheme: A state $\rho_S$ is coupled  to a memory register via unitary $U_{SM}$, followed by \textit{biased}-\textit{unsharp} POVM elements  $E_k$ on the memory register. The memory is reset to its initial state $\rho_M$ using a thermal resource $\rho_B$, before the device is used again.}
		\label{fig:diag}
	\end{figure}

	\begin{figure}[ht!]
		\centering
		\includegraphics[width=80mm]{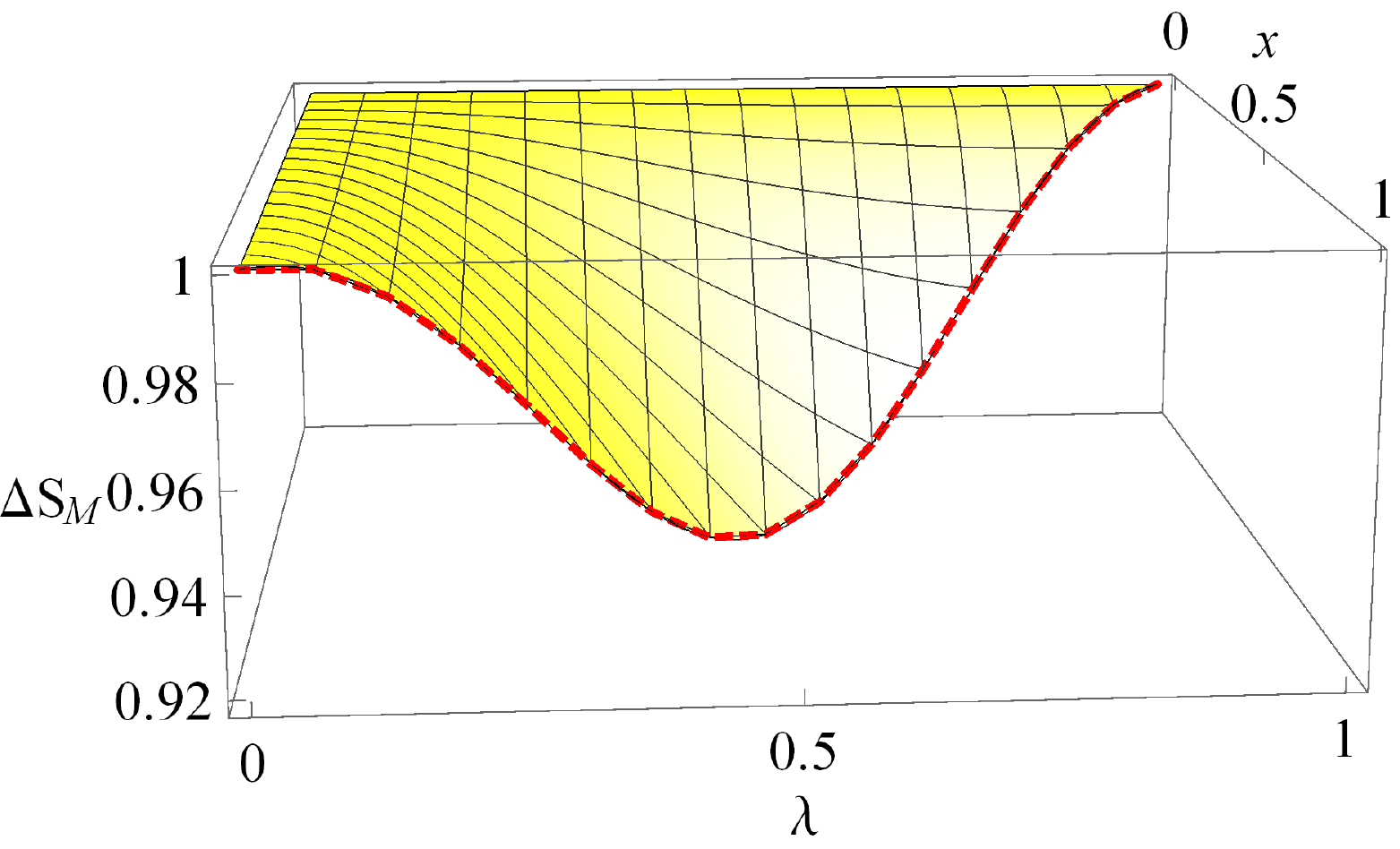}
		\includegraphics[width=80mm]{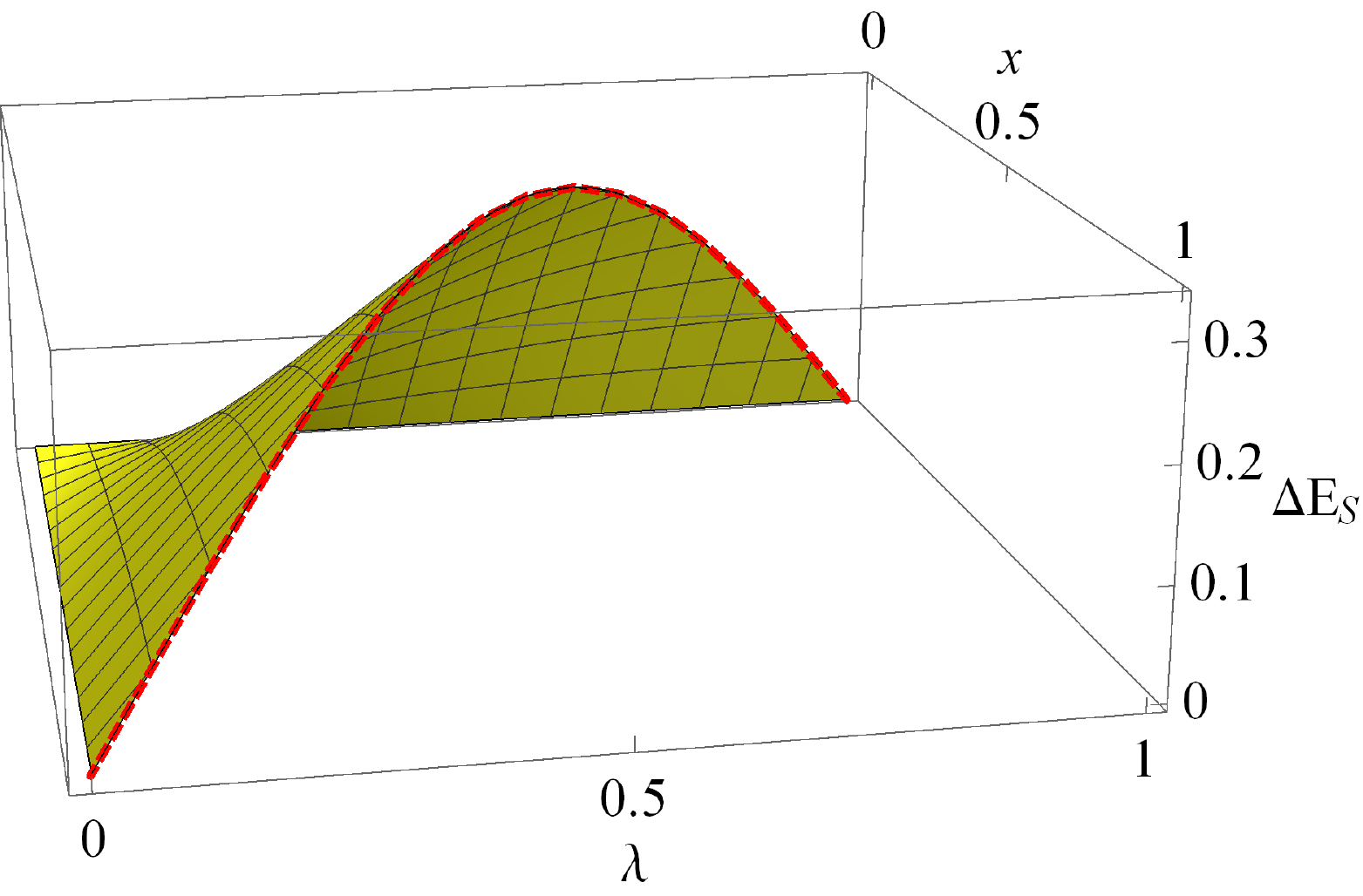}
		\includegraphics[width=80mm]{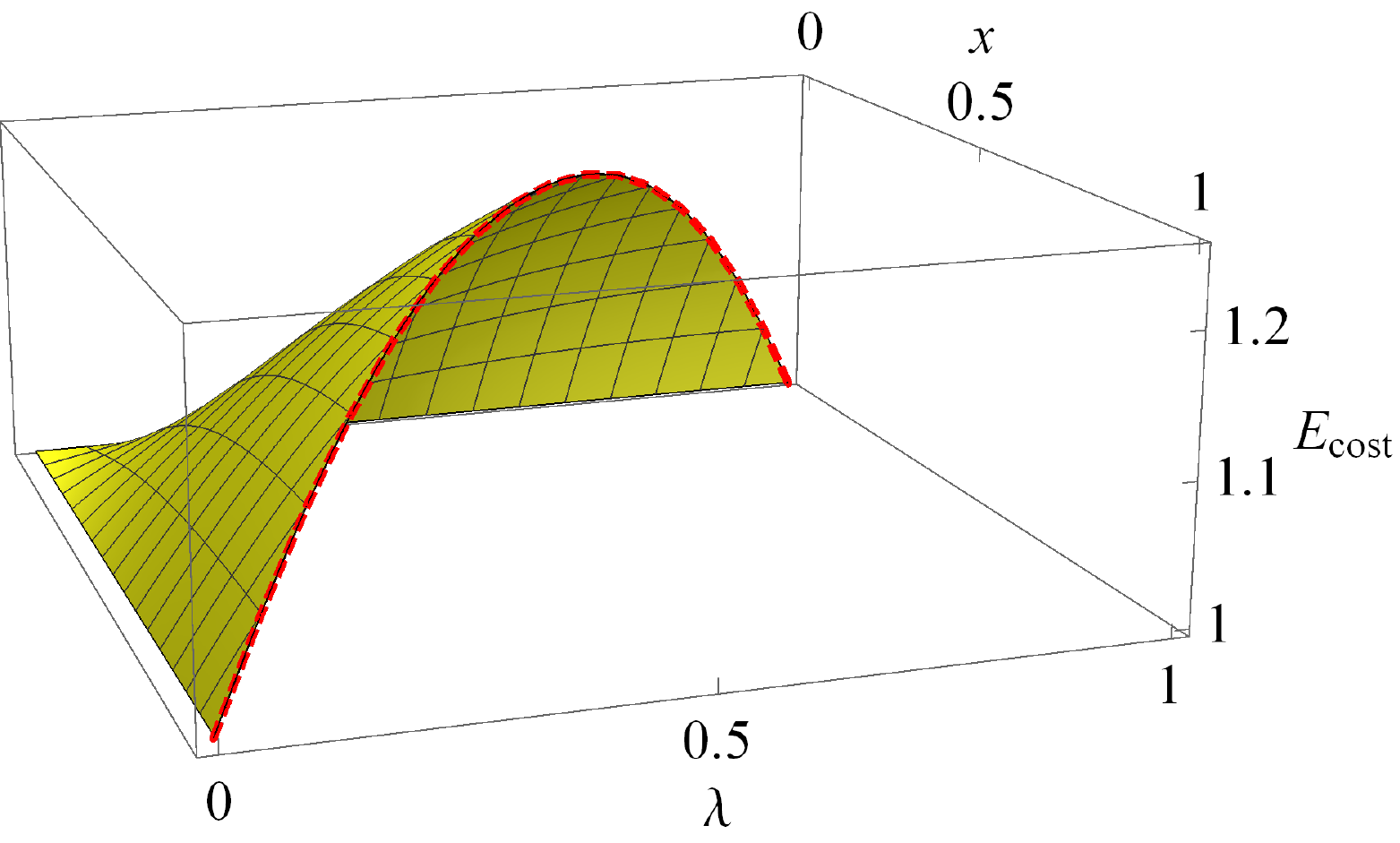}
		\caption{(Color online) Depicting the contributions to the energy cost $E_{cost}$ as defined in Eq. (\ref{eq:Ecost}) with respect to \textit{bias} ($x$) and \textit{sharpness} ($\lambda$) (dimensionless) parameters, such that  $x+\lambda \le 1$, with the red (dashed)  curve pertaining to  equality.}
		\label{fig:EntopyChange}
	\end{figure}

	The register $M$ in the measurement step stores a  measurement outcome $k$ in a state $\rho_{M, k}^\prime \in \mathcal{S} (H_M)$. In order to read out the measurement outcome from the register, one would apply  the projectors $\{\Pi_k \}_{k}$ satisfying $\sum_k \Pi_k = \mathbb{1}$, on the respective subspaces $H_k$. Accordingly, the implementation of a quantum measurement is described by a tuple $(\rho_M, U_{SM}, \{\Pi_k\})$ where $\rho_M$ and $U_{SM}$ denote the initial state of the register and the unitary operator describing the interaction between  system $S$ and  register $M$. Therefore, one can think of the measurement process as  a channel which maps  an input state $\rho_S$  to an output state
	\begin{equation}
		\rho_{SM, k}^\prime = (\mathbb{1} \otimes \Pi_k)  U_{SM} (\rho_S \otimes \rho_M) U_{SM}^\dagger (\mathbb{1} \otimes \Pi_k) / p_k.
	\end{equation}
	with probability $p_k = \opt{Tr} \big[ (\mathbb{1} \otimes \Pi_k)  U_{SM} (\rho_S \otimes \rho_M) U_{SM}^\dagger \big]$, such that $\rho_{SM}^\prime = \sum_{k} p_k~ \rho_{SM, k}^\prime $, and the corresponding reduces states of the system and the memory register are respectively given by $\rho_S^\prime = \opt{Tr}_M[\rho_{SM}^\prime]$ and  $\rho_M^\prime = \opt{Tr}_S[\rho_{SM}^\prime]$.\bigskip

	The total energy cost for measurement  and resetting step turns out to be    \cite{abdelkhalek2016fundamental}
	\begin{equation}\label{eq:Ecost}
		E_{cost} =     \Delta E_{S} +\frac{1}{\beta}  \Delta S_M,
	\end{equation}
	with $\Delta E_S = \opt{Tr}[H_S (\rho_S^\prime - \rho_S)]$ and  $\Delta S_M = S(\rho_M^\prime) - S(\rho_M)$. Thus the total energy cost is essentially determined by the entropy change in the memory.    In what follows, we  bring out the effect of \textit{biased}-\textit{unsharp} measurements on this quantity. Such measurements, often called \textit{inefficient} measurements are characterized by Kraus operators $\{E_i\}$, such that the post-measurement state is given by $\rho^\prime = \sum_i E_i \rho E_i^\dagger$, with $1\le i \le r$, where $r$ is the Kraus \textit{rank}, and  $r=1$ corresponds to  \textit{efficient} measurements.  	Rewriting  Eq. (\ref{eq:Epm})  as
	\begin{equation}\label{eq:EpmNew}
		E_{\pm} = \frac{1 - \lambda \pm x}{2} \mathbb{1} + \lambda \Pi^\pm.
	\end{equation}
	Here, $\lambda = |\vec{m}|$ is the \textit{sharpness} parameter and $\Pi^\pm = (1 \pm \hat{Q})/2$ are the sharp projectors corresponding to observable  $ \hat{Q} = \hat{m}.\vec{\sigma}$,  assumed to be of the form given in Eq. (\ref{measurement}).
	We take a simple model with of the system prepared in  state $\rho_S = \frac{1}{2}|0\rangle \langle 0| + \frac{1}{2} | 1 \rangle \langle 1|$, a statistical mixture of the eigenstates of Hamiltonian $H_S = \omega_S \sigma_z$.  Further, the memory is assumed to  be in a two qubit state $\rho_M = |0\rangle \langle 0|_{M_A} \otimes \frac{1}{2} \mathbb{1}_{M_B}$. With the  projectors $P_k = |k\rangle \langle k| \otimes \frac{1}{2} \mathbb{1}_{M_B}$,   $k=0,1$ and the unitary interaction between the system and memory of the form \cite{abdelkhalek2016fundamental}
	\begin{align}
		U_{SM} &= \Big(|0\rangle \langle 0|_{S} \otimes \mathbb{1}_{M_A} + |1\rangle \langle 1|_{S} \otimes \sigma^x_{M_A} \Big) \otimes \mathbb{1}_{M_B},
	\end{align}
	one can show  that the  measurement device represented by [$U_{SM}, \rho_M, \{P_k\}$]  outputs the correct  state $\rho_{M,k}^\prime = |k\rangle \langle k|_{M_A} \otimes \mathbb{1}_{M_B} $.   However, we are interested in the situation when $\{P_k\}$ are not ideal projective measurements but are \textit{biased} and have some  \textit{unsharp}. This is  incorporated in the above scheme by replacing $|k\rangle \langle k |$ by $E_k$ given in Eq. (\ref{eq:EpmNew}), such that $\{P_k\} \rightarrow \{\tilde{P}_k\} =  \{E_k \otimes \frac{1}{2} \mathbb{1}_{M_B} \}$, we have 
	\begin{align}
		\rho_{SM}^\prime &= \sum_{k = \pm }  (\mathbb{1} \otimes \tilde{P}_k) U_{SM} (\rho_S \otimes \rho_M) U_{SM}^\dagger (\mathbb{1} \otimes \tilde{P}_k).
	\end{align}
	The normalized reduced states of the system and memory are respectively given by 
	 
	\begin{align}
		\rho_S^\prime &= \frac{1}{2} \begin{pmatrix}
			                                                 1 + R_c   &               0 \\
		                                                                 0                    &          1 -  R_c 
	                                                    	\end{pmatrix}, \label{eq:rhoSprime} \\ 
			\rho_M^\prime   &=\begin{pmatrix}
		                     	\frac{1}{4} + \frac{1}{2} R_c         &                            0                                   &        \frac{1}{2} R_s                                  &                         0 \\
				                              0                                        &                 \frac{1}{4} + \frac{1}{2} R_c &             0                                                    &   \frac{1}{2} R_s \\
				               \frac{1}{2} R_s^*            &                             0                                   &                \frac{1}{4} - \frac{1}{2} R_c   &  0 \\
				                              0                                        &                  \frac{1}{2} R_s^*                   &                        0                                         &  \frac{1}{4} -  \frac{1}{2} R_c 
			\end{pmatrix}. \label{eq:rhoMprime}
		\end{align}
 
\begin{figure}[ht!]
		\centering
		\includegraphics[width=80mm]{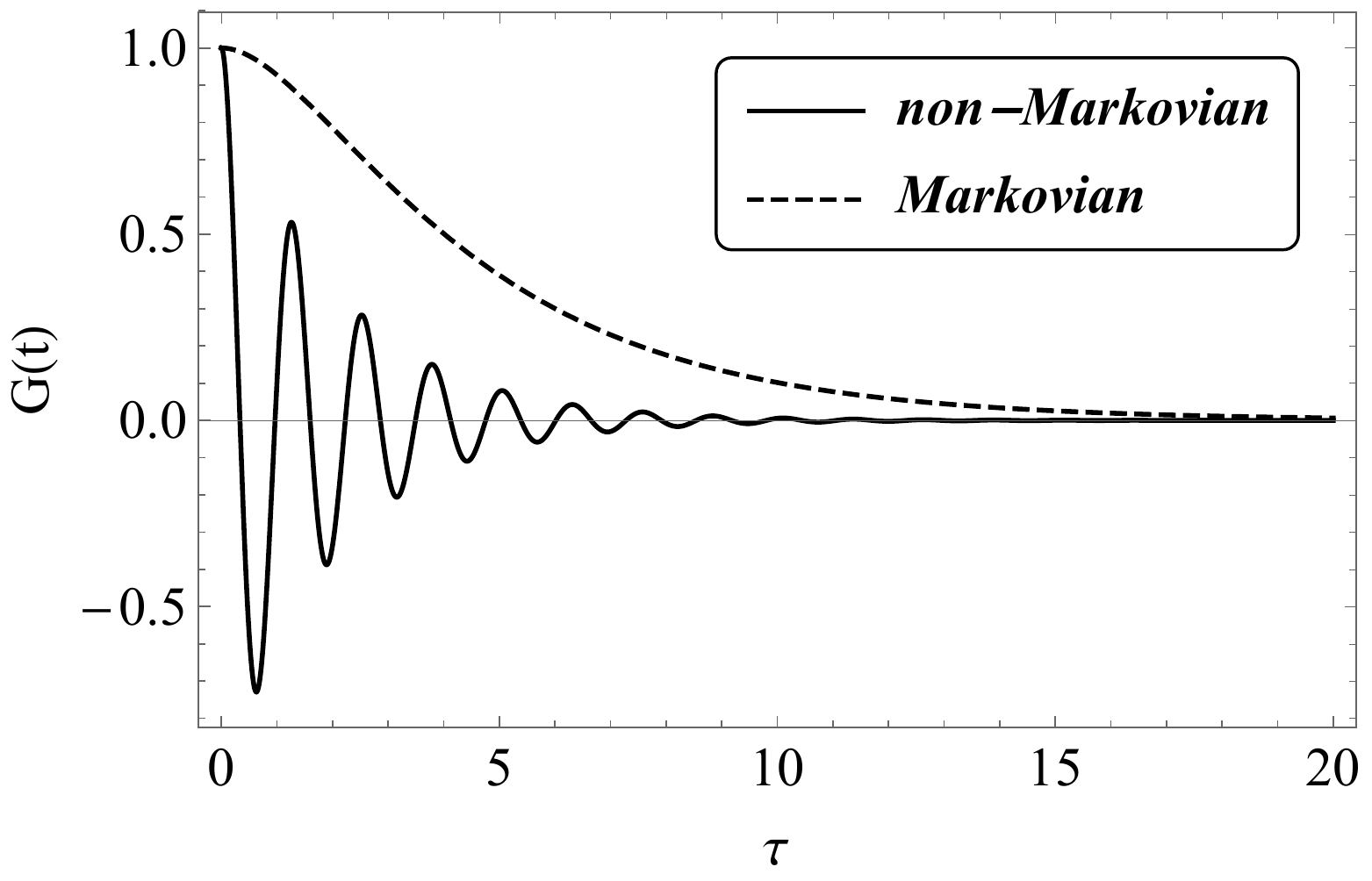}
		\includegraphics[width=80mm]{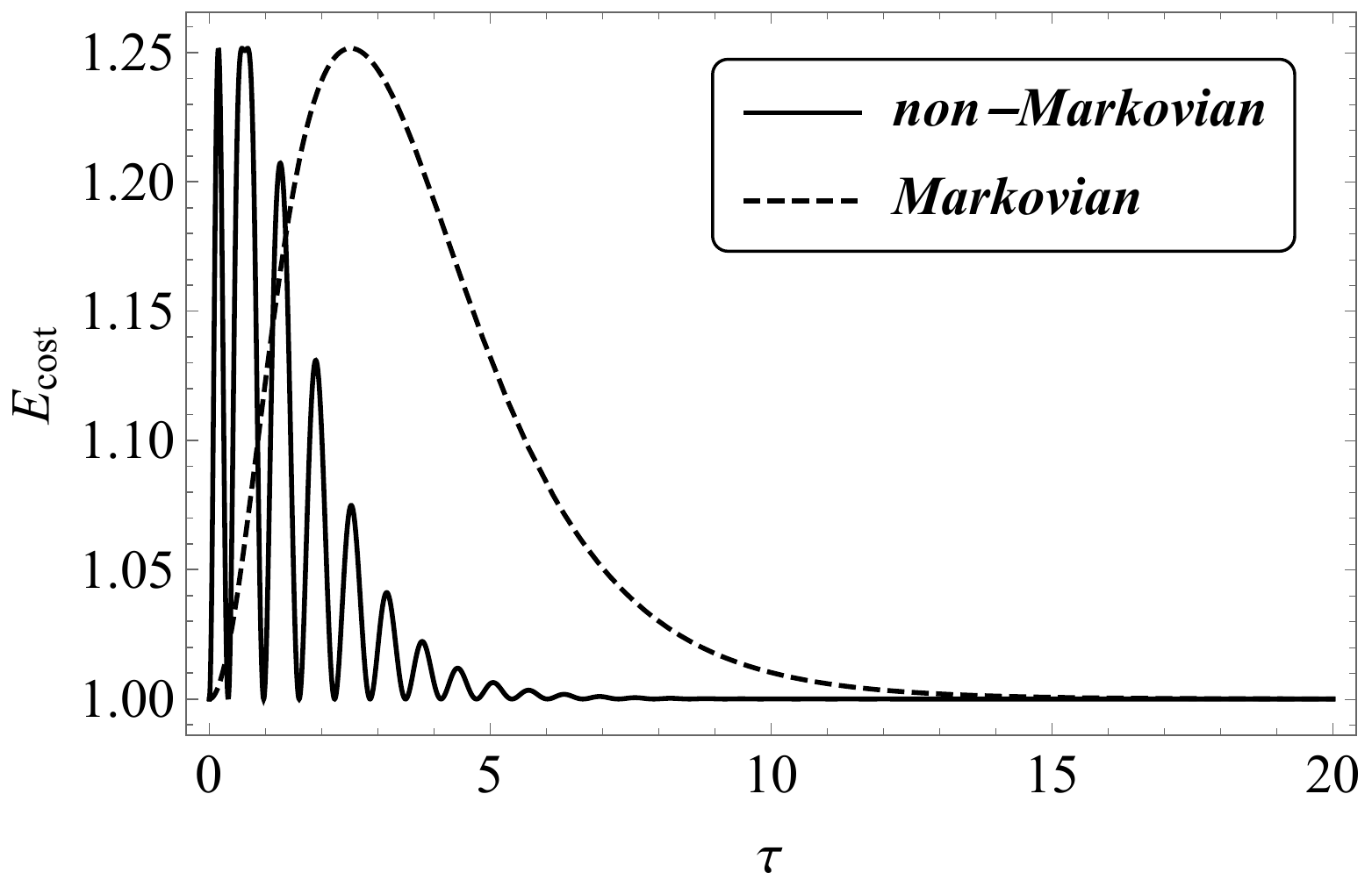}
		\caption{The memorial kernel $G(\uptau)$ (top) and energy cost $E_{cost}$ (bottom) as a function of dimensionless time parameter $\tau$. }
		\label{fig:EcostAD}
	\end{figure}

	Here $R_c = \frac{x \lambda}{1 + x^2 + \lambda^2} \cos(\Theta)$, $R_s = \frac{x \lambda}{1 + x^2 + \lambda^2} e^{i \Phi} \sin(\Theta)$ are introduced for convenience, and $\Theta$ and $\Phi$ are the measurement parameters defined in Eq. (\ref{measurement}). It follows that with either $x=0$ or $\lambda = 0$, both system as well as memory are found to be in maximally mixed state.  The energy cost is proportional to the entropy change $\Delta S_M = S(\rho_M^\prime) - S(\rho_M)$ in the memory 
	\begin{align}\label{eq:EntropyChange}
		\Delta S_M &= - \sum\limits_{n=1}^4  \eta_{n} \log_2(\eta_n) - 1,
	\end{align}
	where $\eta_{n}$  are the eigenvalues of $\rho_M^\prime$, with $\eta_1 = \eta_2 = \frac{1}{4}(1  + \frac{2x \lambda}{1+x^2 + \lambda^2}), \eta_3 = \eta_4 = \frac{1}{4} (1 -  \frac{2x \lambda}{1+x^2 + \lambda^2})$, independent of the measurement parameters $\Theta$ and $\Phi$. This quantity is depicted in Fig. (\ref{fig:EntopyChange}) with respect to \textit{bias} and \textit{sharpness} parameters. A decrease in $\Delta S_M$ is observed for   non-zero values of these parameters, and is found to be minimum for $x=\lambda=1/2$.  Further, the contribution to energy cost due to change in the system state is given by
	\begin{align}\label{eq:DeltaES}
		\Delta E_{S} &= \opt{Tr}\big[ H_S (\rho_{S}^\prime - \rho_S) \big]
		= \frac{2 x \lambda \omega_S}{1 + x^2 + \lambda^2} \cos(\Theta).
	\end{align}
 In this particular example,  the measurement on the system is performed in $\{ \ket{0}, \ket{1}\}$ basis, so  we set $\Theta = \Phi = 0$, which amount to $\mathcal{Q} = \sigma_z$ in Eq. (\ref{measurement}). With this setting. $\Delta E_S$ is depicted in Fig. (\ref{fig:EntopyChange}) for $\omega_S = 1$, and attains maximum value for $x=\lambda = 1/2$. The total energy cost $E_{cost}$ is positive and is also maximum for the measurement characterized by equal \textit{bias} and \textit{sharpness}. One can map this scenario with the non-Markovian amplitude damping channel for which the \textit{bias} and \textit{sharpness} are respectively  given by $x = |G^2(t) - 1|$ and $\lambda = |G(t)|^2$, for  $\Theta = 0$, see Table (\ref{tabBiasUnsharp}), where $G(t)$ is the memory kernel with following form \cite{bylicka2014non}
 
 \begin{align}
 	G(\uptau) = e^{-\uptau/2}\Bigg[\cosh(\sqrt{1-2R}~\uptau/2) + \frac{\sinh(\sqrt{1-2R}~\uptau/2)}{\sqrt{1-2R}} \Bigg].
 \end{align}
	Here, $R$   is proportional to the coupling strength and  $\uptau$, is dimensionless time. The regimes $2R \le 1$ and $2R>1$ correspond to Markovian and non-Markovian dynamics, respectively. The time behavior of the memory kenel is depicted in  Fig. (\ref{fig:EcostAD}), and is found to acquire negative values under  non-Markovian dynamics. As time increases,  $G(\uptau) \rightarrow 0$, and an arbitrary state subjected to AD channel settles to the ground state $\ket{0}$. Correspondingly the \textit{bias} ($x$) and \textit{sharpness} ($\lambda$) parameters tends to $1$ and $0$, respectively, and the  POVM elements in Eq. (\ref{eq:EpmNew}) become $\{E_{+} = \mathbb{1}, E_{-} = 0 \}$. Therefore, the fact that system is eventually found in ground state is equivalent to the statement that the POVM reduces to the identity operation. Notice that $G(\uptau)$ damps quickly as the coupling strength is increased. Therefore, the sharpness of our POVM decreases rapidly with increase in the degree of non-Markovianity of the  noisy channel. This fact is reflected in the energy cost of performing such measurements, as depicted in Fig. (\ref{fig:EcostAD}).
	
	\section{Conclusion}\label{Conclusion}
	Generalized dichotomic measurements characterized by \textit{bias} and \textit{sharpness} provide a way to take into account the different causes which make a measurement non-ideal. The \textit{bias} quantifies tendency of a measurement to favor one state over the other  while as sharpness is proportional  to the  precision of the measurement. In this work, we have shown how the \textit{bias} and \textit{sharpness} change under the action of a dynamical process (\textit{e.g.} quantum channels) from the perspective of  the Heisenberg picture. Specifically, we considered various quantum channels, both Markovian and non-Markovian.  We have shown the unital channel  induce \textit{unbiased} measurements on a qubit state. Also, the conjugate of a unital channel acting on a projective measurement generates an \textit{unbiased} POVM.    Further, Markovian channels are shown to lead to measurements for which sharpness is a monotonically decreasing function of time. Hence, for unital channels, this provides a witness for P-indivisible form of non-Markovian dynamics. \\

	Measurement process is central in carrying out operations with quantum devices in a controlled manner. With increasing complexity of quantum devices, the energy supply for carrying out the elementary quantum operations must be taken into account.  We investigated the effect of \textit{bias} and \textit{sharpness} parameters on the energy cost of the measurement. The energy cost is proportional to  the entropy of  memory register which is found to decrease in presence of \textit{biased}-\textit{unsharp} measurements, however the total energy cost is found to increase under such measurements.

	The present work may be extended to higher dimensional systems and by considering  other definitions of non-Markovianity-- via CP-divisibility of channels.

	\section*{Acknowledgement}
	JN's  work was supported by the project  ``Quantum Optical Technologies" carried out within the International Research Agendas programme of the Foundation for Polish Science co-financed by the European Union under the European Regional Development Fund. JN also thanks The Institute of Mathematical Sciences, Chennai for hosting him during the initial stage of this work.  SB and SG would like to acknowledge the funding:  Interdisciplinary Cyber Physical Systems (ICPS) program of the Department of Science and Technology 	(DST), India, Grant No. DST/ICPS/QuEST/Theme-1/2019/13. 
	
%
\end{document}